\documentclass{PoS}
\usepackage{amsmath}
\usepackage{amsmath,fleqn,multirow}
\usepackage{subfigure}

\title{Review of COMPASS results on transverse-spin effects in SIDIS}

\ShortTitle{Review of COMPASS results on transverse-spin effects in SIDIS}

\author{\speaker{Nour MAKKE}\\
        \textit{On behalf of the COMPASS Collaboration} \\
       Universit\`a degli studi di Trieste, Via A. Valerio 2, 34127 Trieste, Italy \\
       INFN sezione di Trieste,  Padriciano, 99, 34149 Trieste, Italy \\
       E-mail: \email{nour.makke@cern.ch}}

\abstract{The transversity parton distribution remains a poorly known cornerstone in the nucleon spin structure. While the Collins effect in spin asymmetries in Semi-Inclusive DIS (SIDIS) is one crucial tool to address the transversity function, the most promising alternative is the azimuthal asymmetry in SIDIS when a hadron pair is detected in the final state. In this case, the chiral-odd transversity function is coupled to another chiral-odd function, i.e. the hadron-pair interference fragmentation function (IFF). The measurement of azimuthal asymmetries in hadron-pair production on a transversely polarised nucleon target has been performed at COMPASS using a 160 GeV/c muon beam of CERN's M2 beam line. Results from the 2007 and 2010 recent measurements will be presented and compared to model predictions.}

\FullConference{The European Physical Society Conference on High Energy Physics \\
		 18-24 July, 2013\\
		 Stockholm, Sweden}

\begin{document}

\section{Introduction}
The distribution of transversely polarised quarks in a transversely polarised nucleon, denoted by $h_1(x)$ or $\Delta_T q(x)$ was first introduced by Ralston and Soper in 1979 \cite{Ralston:1979ys}. Due to its chiral-odd nature, the transversity is not measurable in inclusive Deep-Inelastic Scattering (DIS). It could be assessed in semi-inclusive DIS of leptons off transversely polarised nucleons by measuring single-spin azimuthal asymmetries (SSA) of cross-sections, where it couples to another chiral-odd function, the spin-dependent fragmentation function. One of the main advantages of SIDIS is that the Collins and Sivers effects, as well as the other transverse momentum dependent (TMD) effects, generate different azimuthal asymmetries and thus do not mix. COMPASS \cite{Ageev:2006da,Alexakhin:2005iw} and HERMES \cite{Airapetian:2004tw} collaborations presented the first data collected with transversely polarised protons and deuterons in 2004 where a clear evidence of transverse SSA's on proton was observed in the covered kinematics.

The model-independent expression for the SIDIS cross-section for transversely polarised target can be written in the following way

\vspace{-0.1cm}
{\footnotesize
\begin{eqnarray}
\label{eq:x_sec_mod}
  && \hspace*{-0.0cm}\frac{{d\sigma }}{{dxdydzdP_{hT}^2d{\varphi _h}d{\varphi _S}}} = \\
 &&\hspace*{0.1cm}  \left[ {\frac{{\cos \theta }}{{1 - {{\sin }^2}\theta {{\sin }^2}{\varphi _S}}}} \right]  \left[ {\frac{\alpha }{{xy{Q^2}}}\frac{{{y^2}}}{{2\left( {1 - \varepsilon } \right)}}\left( {1 + \frac{{{\gamma ^2}}}{{2x}}} \right)} \right]  \left( {{F_{UU,T}} + \varepsilon {F_{UU,L}}} \right) \times \nonumber
\end{eqnarray}
\[\begin{gathered}[h!]\hspace*{-0.0cm} \left( \begin{gathered}
  1 + \cos {\varphi _h}  \sqrt {2\varepsilon \left( {1 + \varepsilon } \right)} A_{UU}^{\cos {\varphi _h}} + \cos {2{\varphi _h}}  \varepsilon  A_{UU}^{\cos {2{\varphi _h}}} +   \lambda \sin {\varphi _h}  \sqrt {2\varepsilon \left( {1 - \varepsilon } \right)} A_{LU}^{\sin {\varphi _h}} +
  \hfill\\
  \frac{{\text{P}}_{\text{T}}}{{{\sqrt {1 - {{\sin }^2}\theta {{\sin }^2}{\varphi _S}} }}} \left[ \begin{gathered}
  \sin {\varphi _S}  \left( \cos \theta {\sqrt {2\varepsilon \left( {1 + \varepsilon } \right)}  A_{UT}^{\sin {\varphi _S}}} \right) +  \hfill \\
  \sin \left( {{\varphi _h} - {\varphi _S}} \right)  \left( {\cos \theta A_{UT}^{\sin \left( {{\varphi _h} - {\varphi _S}} \right)} + \frac{1}{2}\sin \theta\sqrt {2\varepsilon \left( {1 + \varepsilon } \right)} A_{UL}^{\sin {\varphi _h}}} \right) +  \hfill \\
  \sin \left( {{\varphi _h} + {\varphi _S}} \right)  \left(\cos \theta  {\varepsilon A_{UT}^{\sin \left( {{\varphi _h} + {\varphi _S}} \right)} + \frac{1}{2}\sin \theta \sqrt {2\varepsilon \left( {1 + \varepsilon } \right)} A_{UL}^{\sin {\varphi _h}}} \right) +  \hfill \\
  \sin \left( {2{\varphi _h} - {\varphi _S}} \right)  \left( \cos \theta{\sqrt {2\varepsilon \left( {1 + \varepsilon } \right)}  A_{UT}^{\sin \left( {2{\varphi _h} - {\varphi _S}} \right)} + \frac{1}{2}\sin \theta \varepsilon A_{UL}^{\sin 2{\varphi _h}}} \right) +  \hfill \\
  \sin \left( {3{\varphi _h} - {\varphi _S}} \right)  \left(\cos \theta {\varepsilon  A_{UT}^{\sin \left( {3{\varphi _h} - {\varphi _S}} \right)}} \right) +
    \sin \left( {2{\varphi _h} + {\varphi _S}} \right)  \left( {\frac{1}{2}\sin \theta \varepsilon A_{UL}^{\sin 2{\varphi _h}}} \right) \hfill \\
\end{gathered}  \right] +  \hfill \\
  \frac{{{\text{P}}_{\text{T}}}{\lambda}}{{{\sqrt {1 - {{\sin }^2}\theta {{\sin }^2}{\varphi _S}} }}} \left[ \begin{gathered}
  \cos {\varphi _S}  \left( \cos \theta {\sqrt {2\varepsilon \left( {1 - \varepsilon } \right)} A_{LT}^{\cos {\varphi _S}} +  \sin \theta \sqrt {\left( {1 - {\varepsilon ^2}} \right)} {A_{LL}}} \right) +  \hfill \\
  \cos \left( {{\varphi _h} - {\varphi _S}} \right)  \left( \cos \theta {\sqrt {\left( {1 - {\varepsilon ^2}} \right)} A_{LT}^{\cos \left( {{\varphi _h} - {\varphi _S}} \right)} + \frac{1}{2} \sin \theta \sqrt {2\varepsilon \left( {1 - \varepsilon } \right)} A_{LL}^{\cos {\varphi _h}}} \right) +  \hfill \\
    \cos \left( {2{\varphi _h} - {\varphi _S}} \right)  \left( \cos \theta {\sqrt {2\varepsilon \left( {1 - \varepsilon } \right)} A_{LT}^{\cos \left( {2{\varphi _h} - {\varphi _S}} \right)}} \right) + \hfill \\
  \cos \left( {{\varphi _h} + {\varphi _S}} \right)  \left( {\frac{1}{2}\sin \theta \sqrt {2\varepsilon \left( {1 - \varepsilon } \right)} A_{LL}^{\cos {\varphi _h}}} \right) \hfill
\end{gathered}  \right] \hfill
\end{gathered}  \right) \hfill
\end{gathered} \]
}

Here $\phi_{h}$ is the angle of the transverse momentum of the outgoing hadron and $\phi_{S}$ is the azimuthal angle of the quark spin before the hard scattering. The first and second subscripts indicate the beam and the target polarisations ($U$ unpolarised, $L$ longitudinal and $T$ transverse). A total of eight TMD distribution functions are needed to fully describe the transverse-spin and transverse-momentum structure of the nucleon, leading to eight azimuthal asymmetries in Eq.\ref{eq:x_sec_mod}: five Single-Spin (SSA) and three Double-Spin (DSA) target transverse spin-dependent asymmetries.  Because of the smallness of the $\sin\theta$ In the COMPASS kinematics, the impact of the additional terms, represented by $\sin\theta$-scaled longitudinal spin amplitudes and $\theta$-angle dependent factors, is sizeable only in the case of the $A_{LT}^{\cos\phi}$ DSA which remains sizeably affected by the large $A_{LL}$ amplitude. 

\section{Transverse asymmetries in single hadron production}

Among these azimuthal asymmetries, the Collins and Sivers are of particular interest. The collins asymmetries are generated by the Collins effect in the single hadron production. In this mechanism, the Collins fragmentation function $\Delta_T^0 D_q^h$, which describes the correlation between the fragmenting quark spin and the momentum of the produced hadron, introduces a left-right symmetry in the distributions of hadrons reflected in the hadron yields. The Sivers function $\Delta_0^Tq$ (or $f_1^q$) arises from a correlation between the transverse momentum of an unpolarised quark in a transversely polarised nucleon and the nucleon spin. At LO, the asymmetries can be written as

\begin{equation}
 \qquad \quad \quad A _{Coll} = \frac{\sum_q e_q^2 \cdot \Delta_Tq \cdot \Delta_T^0D_q^h}{\sum_0 e_q^2  \cdot q \cdot D_q^h}, \qquad 
A_{Siv} = \frac{\sum_q e_q^2 \cdot \Delta_0^Tq \cdot D_q^h}{\sum_q e_q^2 \cdot q \cdot D_q^h}
\label{C&S}
\end{equation}

The asymmetries are built by comparing the azimuthal distributions of hadrons produced in semi-inclusive DIS on transversely polarised nucleons with opposite spin configurations, in the relevant azimuthal angle and then fitted with the unbind maximum-likelihood method, based on maximum-likelihood fits with the data unbind in $\phi_h$ and $\phi_S$. more details on the analysis can be found in \cite{Adolph:2012sn} (and references therein). COMPASS has collected SIDIS data on transversely polarised target with $^6$LiD (2002-04 years) and NH$_3$ (2007-2010 years), with a target polarisation factor of the order of $50\%$ and $90\%$ and a dilution factor of the order of $0.38$ and $0.15$ respectively. The kinematic range of the measurement is defined by $Q^2 > 1$ (GeV)$^2$, $0.1 < y < 0.9$, $0.003 < x < 0.7$ and $W > 5$ GeV and $z > 0.2$.

\begin{figure}[htbp]
\begin{center}
\includegraphics[height=9.cm,width=.7\textwidth]{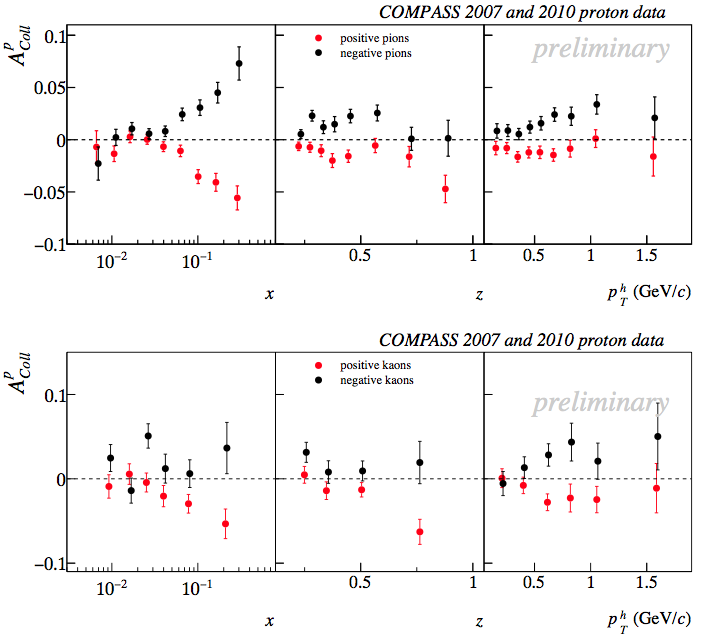}
\caption{The Collins asymmetries for pions (top) and kaons (bottom) measured using a transversely polarised proton target at COMPASS.}
\label{CProt}
\end{center}
\end{figure}

\begin{figure}[htbp]
\vspace{-0.5cm}
\begin{center}
\includegraphics[height=10.cm,width=.7\textwidth]{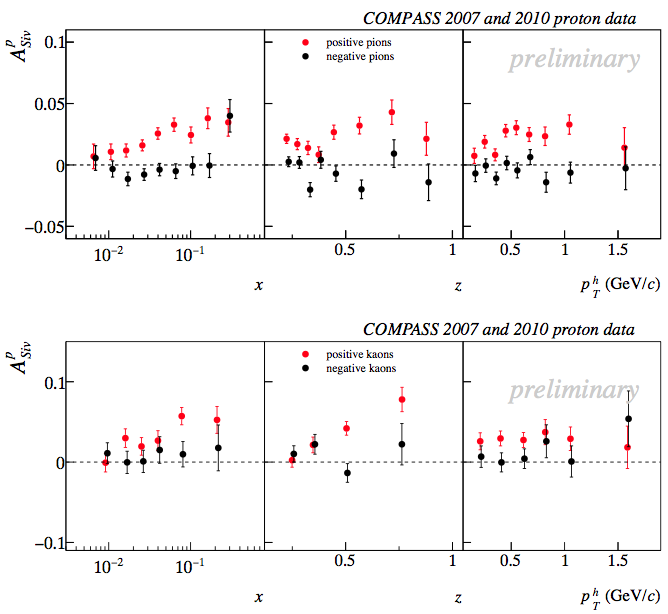}
\caption{The Sivers asymmetries for pions (top) and kaons (bottom) measured using a transversely polarised proton target at COMPASS.}
\label{SProt}
\end{center}
\end{figure}

The analysis of Collins and Sivers asymmetries for charged pions and kaons from the full data set on proton target (data collected in 2007 and 2010) has been recently achieved and the results are shown as a function of the Bjorken variable $x$, the hadron fractional energy $z$ and the hadron transverse momentum $p^{h}_{T}$ in Figs.\ref{CProt} and \ref{SProt}.

\paragraph{Collins asymmetries:} 
The pion asymmetries, which are very similar to the published unidentified hadron ones \cite{Alekseev:2010rw,Adolph:2012sn}, show a zero signal in the sea region and a significant non-zero signal in the valence region with opposite sign for $\pi^+$ and $\pi^-$. This intuitively indicates that the Collins fragmentation functions have the same magnitude and opposite sign \cite{Barone:2010zz} since 
$$A_{Coll,p}^{\pi^+} = e_{u}^2h_{1}^{u}H_1^{\perp,fav} + e_{d}^2h_{1}^{d}H_1^{\perp,unf}, \qquad  A_{Coll,p}^{\pi^-}=e_{u}^2h_{1}^{u}H_1^{\perp,unf} + e_{d}^2h_{1}^dH_1^{\perp,fav}$$
Neglecting the $d$ quark contribution because of its small charge weight,  $|A_{Coll,p}^{\pi^+}| \simeq |A_{Coll,p}^{\pi^-}|$ implies that $H_{\perp}^{fav} = -H_{\perp}^{unf}$.
In the kaon case, the asymmetry show a similar trend, although affected by large statistical uncertainties, in particular for K$^+$ which shows a significant trend towards negative values at high $x$. 
These results surprisingly agree with the existing measurements by the HERMES experiment \cite{Airapetian:2009ae} which has a different kinematic coverage than COMPASS, i.e. the COMPASS $Q^2$ is larger by a factor 2-3 wrt HERMES one in the last $x$ bin.  The same measurement have been performed using a deuterium target and  published in \cite{Ageev:2006da, Alekseev:2008aa} where all asymmetries were found to be compatible with zero. This is intuitively interpreted as a cancellation between the $u$ and $d$ quark contribution in an isoscalar target. Since $A_{Coll,d}^{\pi^+} = A_{Coll,d}^{\pi^-} \simeq (h_1^u + h_1^d)(e_{u}^{2}H_{1}^{\perp,fav} + e_{d}^{2}H_{1}^{\perp,unf})$, we conclude that the transversity $h_1^u$ and $h_1^d$ have the same size and opposite sign. 

\paragraph{Sivers asymmetries}
The Sivers asymmetries for negative pions and kaons are compatible with zero, in contrast with positive pions and kaon asymmetries where a significant signal extends the overall measured $x$ range and increases with $z$. The K$^+$ asymmetry is intriguinly larger than $\pi^+$ asymmetry. This suggests a rather significant role of the sea quarks. At variance with the Collins case, the COMPASS Sivers asymmetries are smaller than HERMES's ones \cite{Airapetian:2004tw} by a factor of two, giving insights of the $Q^2$ evolution of the Transverse Momentum Dependent (TMD) functions. In the case of a deuterium target, all measured Sivers asymmetries are compatible with zero. This is naively interpreted, in the framework of the parton model, as due to opposite Sivers functions for $u$ and $d$ quarks. 

\begin{figure}[htbp]
\begin{center}
\includegraphics[height=5.cm,width=.85\textwidth]{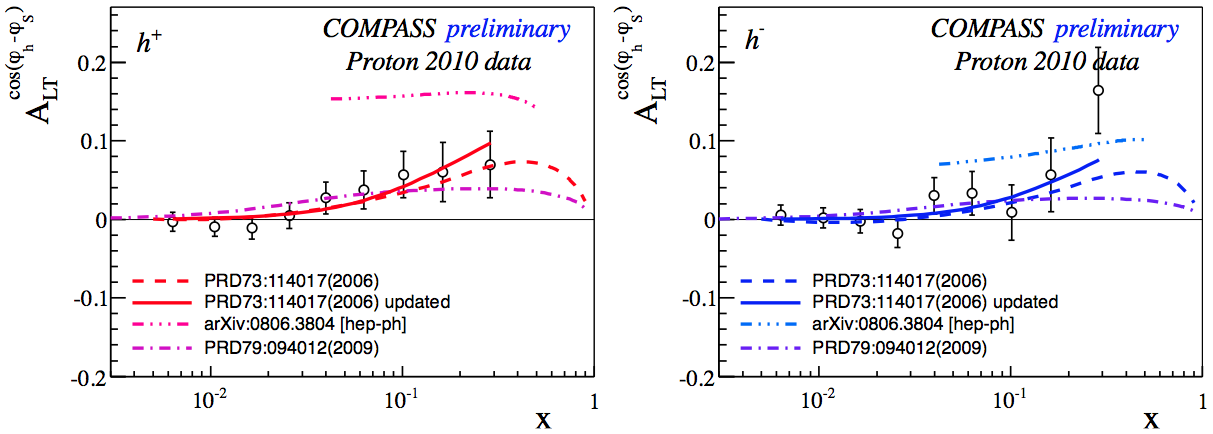}
\caption{"Beyond Collins and Sivers" asymmetries $A-{LT}^{\cos(\phi_h-\phi_S)}$ and $A_{UT}^{\sin\phi_S}$ recently measured at COMPASS using a transversely polarised proton target..}
\label{BCS}
\end{center}
\end{figure}

\paragraph{Other 6 asymmetries}
The other six "beyond Collins and Sivers" asymmetries ($A_{LT}^{\cos(\phi_{h}-\phi_{S})}$, $A_{UT}^{\sin\phi_{S}}$, $A_{UT}^{\sin(3\phi_h -\phi_S)}$, $A_{UT}^{\sin(2\phi_h-\phi_S)}$, $A_{LT}^{\cos\phi_h}$, $A_{LT}^{\cos(2\phi_h-\phi_S)}$) have been also measured by COMPASS on deuterium and proton targets for charged unidentified hadrons as function of $x$, $z$ and $p^{h}_{T}$. These asymmetries are consistent with zero within the statistical accuracy except for $A_{LT}^{\cos(\phi_{h}-\phi_{S})}$ and $A_{UT}^{\sin\phi_{S}}$ where there is an evidence of a non-zero signal. Fig.\ref{BCS} shows the $A_{LT}^{\cos(\phi_{h}-\phi_{S})}$ compared to theoretical predictions from \cite{Kotzinian:2006dw}, \cite{Boffi:2009sh} and \cite{Kotzinian:2008fe} demonstrating a good level of agreement between theory and experimental measurements within the reached statistical accuracy. 

\vspace{0.2cm}

\section{Transverse asymmetries in hadron pair production}

\begin{figure}[htbp]
\begin{center}
\includegraphics[height=5.cm,width=.5\textwidth]{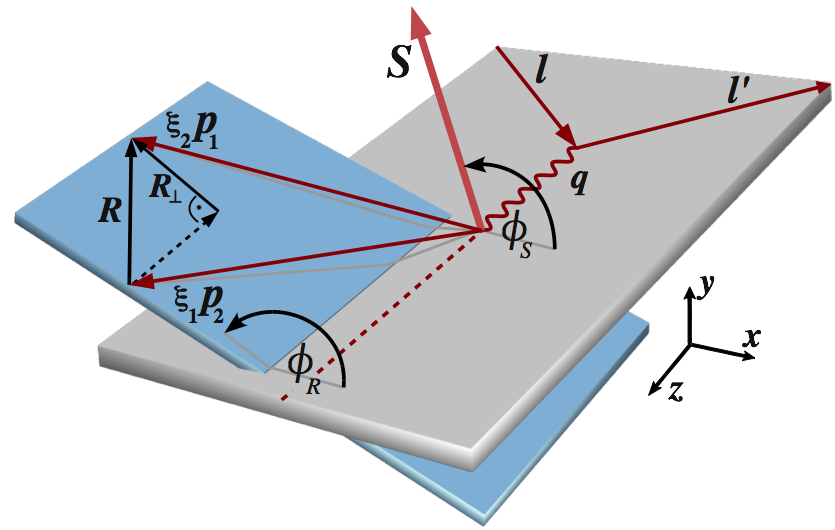}
\caption{Simplified scheme of the dihadron production mechanism: The incoming and the scattered leptons with their 3-momenta \textit{\textbf{l}} and \textit{\textbf{l'}} define the scattering plane (gray). $\phi_{S}$ is the azimuthal angle of the fragmenting quark spin \textbf{S}. The 3-momenta of both hadrons define the lepton plane (blue). The corresponding $\xi_{i}$ values are used in the normalisation of the difference vector \textit{\textbf{R}}, i.e. $\textit{\textbf{R}} = (z_2\textbf{p}_1 - z_1\textbf{p}_2)/(z_1+z_2)=\xi_2\textbf{p}_1 - \xi_1\textbf{p}_2$. $\phi_{R}$ is the azimuthal angle of \textit{\textbf{R}} and \textit{\textbf{R}}$_T$ is its component perpendicular to the 3-momentum of the virtual photon.}
\label{2hSIDIS}
\end{center}
\end{figure}

An alternative approach to assess the transversity function is in dihadron production in SIDIS (Fig. \ref{2hSIDIS}). The dihadron asymmetries of unidentified $h^+h^-$ and identified ($\pi^+\pi^-$, $\pi^+K^-$, $K^+\pi^-$, $K^+K^-$) hadron pairs were measured using the full data set collected using a deuterium ($^6$LiD) target during 2002-04 and on a proton (NH$_3$) target during 2007 and 2010. The measurement is done in the same kinematic range however there additional cuts are applied in this case: each of the selected hadron pairs has to have $z>0.1$ and $x_{F}>0.1$ to avoid any contamination from the target fragmentation region. Exclusively produced $\rho^0$ mesons are rejected by a cut on the missing energy in the reaction, i.e. $E_{miss}>3$ GeV. Finally a cut on $R_{T} > 0.07$ GeV ensures a well defined azimuthal angle $\phi_R$. The full statistics collected using a proton target consists of $45.5 \cdot 10^6$ $h^+h^-$ pairs, of which 28 $\cdot 10^6$ are identified as pion pairs. The deuteron sample consists of $5.8 \cdot 10^6$ $h^+h^-$.

\begin{figure}[htbp]
\begin{center}
\subfigure[2002-04 (top) and 2007 (bottom) $h^+h^-$ pairs asymmetries measured using a transversely polarised deuteron and proton targets respectively compared with model productions \cite{Bacchetta:2006un,She:2007ht}.]{\label{2h020407}\includegraphics[height=6.cm,width=.75\textwidth]{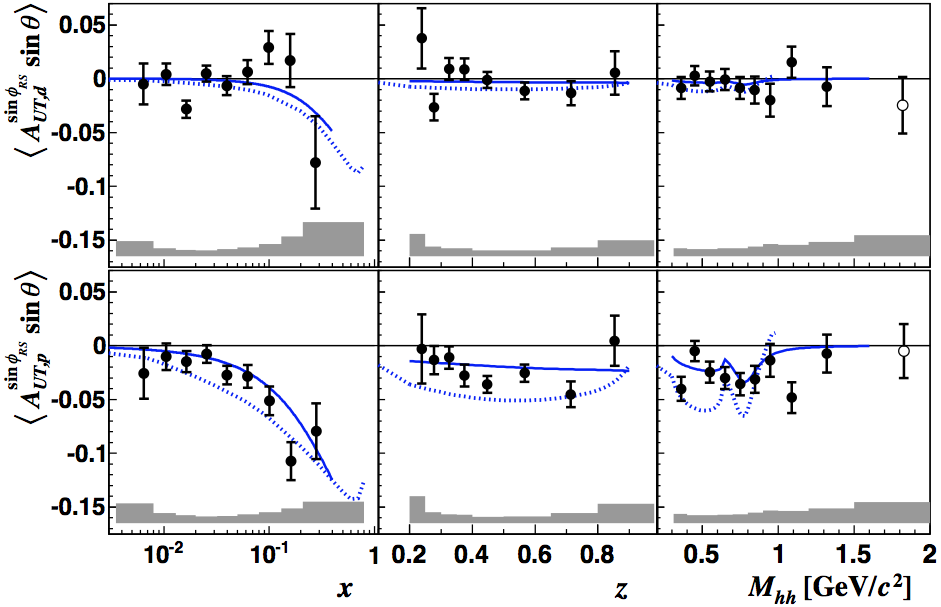}}
\subfigure[$\pi^+\pi^-$ asymmetries from combined 2007 and 2010 proton data in comparison with HERMES data \cite{Airapetian:2008aa} and model predictions \cite{Bacchetta:2006un,She:2007ht} in the valence region ($x > 0.032$).]{\label{2pi}\includegraphics[height=4.cm,width=.75\textwidth]{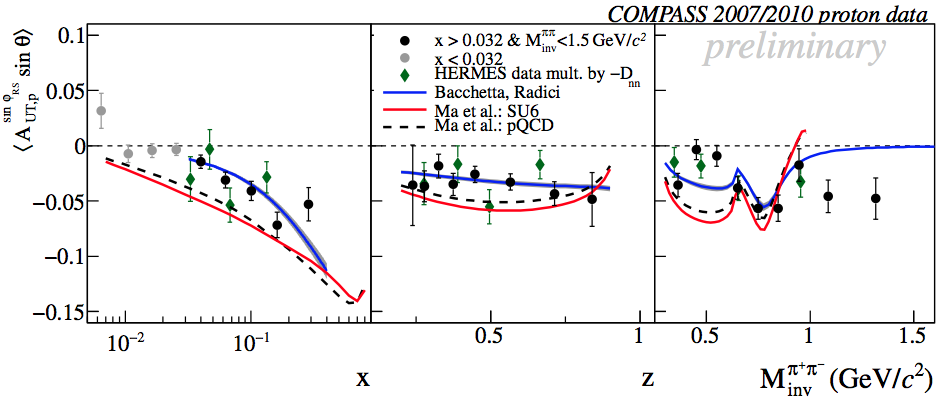}}
\caption{Hadron pair asymmetries on transversely polarised deuteron and proton targets at COMPASS.}
\label{}
\end{center}
\end{figure}

\begin{figure}[htbp]
\begin{center}
\includegraphics[height=11.cm,width=.85\textwidth]{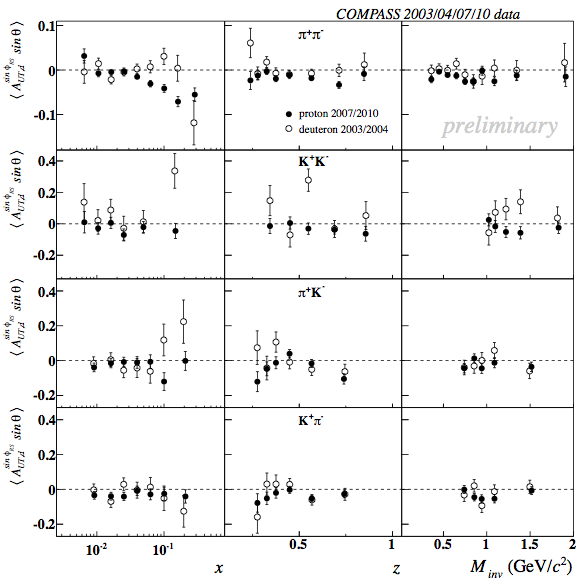}
\caption{New Hadron pair asymmetries measured at COMPASS using a transversely polarised deuteron (2003/04) and proton (2007/10) targets using the RICH for particle identification in the range [3,10 GeV].}
\label{2h_all}
\end{center}
\end{figure}

Figs. \ref{2h020407}(top) and \ref{2h_all} show respectively the asymmetries for unidentified and identified hadron pair asymmetries measured using an isoscalar target. The asymmetries are found to be consistent with zero within the statistical accuracy. Furthermore no specific trend is visible in any of the $x$, $z$ or $M_{inv}$ dependences, in agreement with the existing theoretical models predictions and in line with the COMPASS measurement of the Collins asymmetry on deuterium. This observation can be interpreted as an almost complete cancellation between the up ($h^u_1$) and down ($h^d_1$) quark transversity on isoscalar target which is predicted by the available theoretical models \cite{Bacchetta:2006un,She:2007ht}. \\
The first COMPASS measurement of the dihadron asymmetry of $h^+h^-$ pairs on a proton target was performed using the data collected in 2007 and shown in Fig.\ref{2h020407} (bottom) as a function of $x$, $z$ and $M_{inv}$. While no specific trend is visible versus $z$, a large asymmetry up to -10\% is measured in the valence $x$-region and a significant negative signal versus $M_{inv}$ is measured around the $\rho^{0}$ mass of 0.77 (GeV/c)$^2$. In order to improve the results in terms of statistics and to allow further analyses, the full beam time in 2010 was dedicated to collect data again on a transversely polarised target. The asymmetry measurement based on 2010 data set showed a very good agreement and consistency with the results from 2007 and therefore a recently new analysis was performed by combining both 2007 and 2010 data sets. COMPASS has the particle identification facility using a Ring Imaging Cherenkov detector and ensures a very good separation of the particle types in the momentum range [3 GeV,50 GeV]. 
The resulting asymmetries for identified dihadron are shown in Fig.\ref{2h_all}. The pion pair asymmetry shows a significant negative signal up to -6\% in the valence $x$-region and a pronounced peak around the $\rho^0$ mass in it's $M_{inv}$ dependence. The kaon pair asymmetry is mostly dominated by large statistical uncertainties and is compatible with zero in almost all kinematic bins except at large $M_{inv}$. For mixed pairs ($\pi^+K^-$ and $K^+\pi^-$), the asymmetries are consistent with zero. 

A comparison between COMPASS $\pi^+\pi^-$ asymmetry and existing measurement by the HERMES experiment \cite{Airapetian:2008aa} and theoretical model predictions by Bacchetta \textit{et al.} \cite{Bacchetta:2006un}  and Ma \textit{et al.} \cite{She:2007ht} is shown in Fig.\ref{2pi}. COMPASS and HERMES measured $\pi^+\pi^-$ asymmetries agree in the commonly covered $x$ range despite the different $Q^2$ coverage in each $x$ bin. The experimental trend in $x$ and the observed peak around the $\rho^0$ mass are well reproduced in both model predictions. However, the agreement in $z$ and the remaining mass regions is rather poor.

\section{Conclusion}

An overview of COMPASS investigation on transverse-spin effects and transverse momentum-dependent effects in SIDIS has been presented. A full set of recent results on the eight azimuthal asymmetries, of pions and kaons, measured by scattering 160 GeV muon beam off a transversely polarised target in 2007 and 2010 were presented and discussed. While the results of collins and Sivers on deuteron are compatible with zero, a non-zero effect is observed when using a proton target. This observation is deeper investigated via a multi-dimensional analysis of the full data set. A significant signal is also observed in dihadron production, highlighting a non-zero transversity function for up and down quarks.

\end{document}